\documentclass[12pt,prd,showpacs,tightenlines,nofootinbib]{revtex4}
\usepackage{bm}
\usepackage{graphics}
\usepackage{rotating}
\usepackage{epsfig}
\begin{document}
\title{\begin{flushright}{\rm\normalsize HU-EP-05/73}\end{flushright}
Masses and electroweak properties of light mesons in
the relativistic quark model}  
\author{D. Ebert}
\affiliation{Institut f\"ur Physik, Humboldt--Universit\"at zu Berlin,
Newtonstr. 15, D-12489  Berlin, Germany}
\author{R. N. Faustov}
\author{V. O. Galkin}
\affiliation{Institut f\"ur Physik, Humboldt--Universit\"at zu Berlin,
Newtonstr. 15, D-12489 Berlin, Germany}
\affiliation{Dorodnicyn Computing Centre, Russian Academy of Sciences,
  Vavilov Str. 40, 119991 Moscow, Russia}

\begin{abstract}
The masses, pseudoscalar and vector weak decay constants
and electromagnetic form factors of light $S$-wave mesons
are studied in the framework of
the relativistic quark model based on the quasipotential approach. We
use the same model assumptions and parameters as in our
previous investigations of heavy meson and baryon properties. The
masses and wave functions of the ground state and radially excited
$\pi$, $\rho$, $K$, $K^*$ and $\phi$ mesons, obtained by solving
numerically the relativistic Schr\"odinger-like equation with the
complete relativistic $q\bar q$ potential including both
spin-independent and spin-dependent terms, are presented. 
Novel relativistic expressions 
for the weak decay constants of the pseudoscalar and vector mesons are
derived. It is shown that the intermediate negative-energy quark
states give significant contributions which essentially decrease the decay
constants bringing them in agreement with experimental data.    
The electromagnetic form factors of the pion, charged and neutral kaon
are calculated in a broad range of the space-like momentum
transfer. The corresponding charge radii are determined. All results
agree well with available experimental data. 
\end{abstract}


\maketitle

\section{Introduction}
\label{sec:int}
The theoretical investigation of the properties of light mesons such
as  $\pi$, $\rho$, $K$, $K^*$ and $\phi$ is a longstanding  problem
which plays an important role in understanding the low-energy
QCD. The description of these mesons   within the constituent quark 
model presents additional difficulties compared to heavy-light mesons
and heavy quarkonia. In fact, due to the highly relativistic dynamics 
of light quarks, the $v/c$ and $1/m_q$ expansions are completely
inapplicable in 
the case of light mesons and the QCD coupling constant $\alpha_s$ at
the related scale $\mu$ is rather
large. Moreover, the behaviour of $\alpha_s(\mu^2)$ in
the infrared region is unknown and thus model dependent (exhibiting, e.g.,
freezing behaviour, etc.). The pseudoscalar mesons $\pi$ and $K$
produce a special problem, 
since their small masses originate from their Goldstone nature caused
by the broken chiral symmetry. 
Therefore the reliable description of light mesons requires the completely
relativistic approach.  It is well known that in the relativistic
studies an important role is played by the Lorentz properties of the
confining quark-antiquark interaction. The comparison of theoretical
predictions with experimental data can provide a valuable information on
the form of the confining potential. Such information is of great
practical interest, because at present it is not possible to obtain the
relativistic $q\bar q$ potential in the whole  range of distances from
the basic 
principles of QCD.\footnote{Recent calculations of the
  nonperturbative $q\bar q$ potential in continuum Yang-Mills theory
  in Coulomb gauge can be found in Ref.~\cite{fr}.} 
Most of the main characteristics of light mesons
are formed in the infrared (nonperturbative) region, thus providing
an important insight in the low-energy properties of $q\bar
q$ interaction. Thus investigation of both static (e.g., masses and
decay constants) and dynamic (e.g., electroweak decay form factors)
properties is of a significant importance.   

Many different theoretical approaches have been used for studying
light mesons, which are based on the
relativized quark model \cite{gi}, the
Dyson-Schwinger and Bethe-Salpeter equations \cite{mr,k}, chiral quark
models with spontaneous symmetry breaking (e.g. the Nambu-Jona-Lasinio
model) \cite{erv}, the relativistic Hamilton dynamics \cite{kt,hjd},
the finite-energy \cite{kp} and light-cone \cite{bkm} sum rules  and 
lattice QCD \cite{ak}.  
Here we consider the possibility of investigating light mesons on the
basis of the three-dimensional relativistic wave equation with the QCD
motivated potential. Our relativistic quark model was originally
constructed for the investigation of hadrons with heavy quarks. It was
successfully applied for the calculation of their masses and various electroweak
decays \cite{efg,egf,fgm,dhb,hbm}. In these studies the 
heavy quark expansion has been used to simplify calculations. We determined
all parameters of our model from few experimental observables (some
masses and decay rates) and keep them fixed in all our subsequent
calculations, thus ensuring its universality.     
While describing the properties of heavy-light 
mesons \cite{egf}, we treated the light quarks in a completely relativistic
way. Recently this approach was applied for calculating the  masses of light
mesons \cite{lmm} and light diquarks inside the heavy baryons
\cite{hbm}. Due to the phenomenological character of our model we
cannot reveal 
the origin of the chiral symmetry breaking and thus the model cannot
describe the chiral limit and the Goldstone nature of the pion. We
consider the pion as the purely bound state of the quark and antiquark
with fixed constituent masses.  

In this paper we extend our previous studies of light mesons and
describe their electroweak properties such as the weak decay
constants and electromagnetic form factors. 
The investigation of decay constants and form factors is an
important issue since it provides a very sensitive test of the light
meson wave functions and, thus, of the quark dynamics in a meson. It
requires the completely relativistic 
consideration of the corresponding decay processes including account
for the 
relativistic transformation of the meson wave functions. The
comparison with the available large set of experimental data tests the
model 
predictions in a broad momentum range and helps to discriminate
between different model assumptions.

The paper is organized as follows. In Sec.~\ref{sec:rqm} we briefly
describe our relativistic quark model, formulate our main assumptions
and give the values of  parameters. Then in Sec.~\ref{sec:lmm} we
present our results for the light meson masses \cite{lmm},
for selfconsistency.  There the procedure of constructing 
the completely relativistic local potential of the light quark
interaction in a meson is described. The obtained
potential is applied for calculating the light $S$-wave meson
masses and wave functions. In Sec.~\ref{sec:dc} the novel relativistic
expressions for the weak decay constants of pseudoscalar and vector
meson are derived. Special attention is paid to including all possible
intermediate quark states. It is argued that the negative-energy
contributions play an essential role. The calculated decay
constants are compared with other predictions and experimental data.  
The electromagnetic form factors of pseudoscalar mesons are studied in
Sec.~\ref{em}. The relativistic expressions for these form factors are
obtained which take into account the contributions of negative-energy quark
states and relativistic transformations of the meson wave functions
from the rest frame to the moving one. The calculated form factors are
plotted in comparison with experimental data. The charged radii of the
pion, charged and neutral kaon are also determined. Our conclusions are
given in Sec.~\ref{sec:concl}.     

\section{Relativistic quark model}
\label{sec:rqm}

  In the quasipotential approach a meson is described by the wave
function of the bound quark-antiquark state, which satisfies the
quasipotential equation  of the Schr\"odinger type~\cite{efg}
\begin{equation}
\label{quas}
{\left(\frac{b^2(M)}{2\mu_{R}}-\frac{{\bf
p}^2}{2\mu_{R}}\right)\Psi_{M}({\bf p})} =\int\frac{d^3 q}{(2\pi)^3}
 V({\bf p,q};M)\Psi_{M}({\bf q}),
\end{equation}
where the relativistic reduced mass is
\begin{equation}
\mu_{R}=\frac{E_1E_2}{E_1+E_2}=\frac{M^4-(m^2_1-m^2_2)^2}{4M^3},
\end{equation}
and $E_1$, $E_2$ are given by
\begin{equation}
\label{ee}
E_1=\frac{M^2-m_2^2+m_1^2}{2M}, \quad E_2=\frac{M^2-m_1^2+m_2^2}{2M}.
\end{equation}
Here $M=E_1+E_2$ is the meson mass, $m_{1,2}$ are the quark masses,
and ${\bf p}$ is their relative momentum.  
In the center-of-mass system the relative momentum squared on mass shell 
reads
\begin{equation}
{b^2(M) }
=\frac{[M^2-(m_1+m_2)^2][M^2-(m_1-m_2)^2]}{4M^2}.
\end{equation}

The kernel 
$V({\bf p,q};M)$ in Eq.~(\ref{quas}) is the quasipotential operator of
the quark-antiquark interaction. It is constructed with the help of the
off-mass-shell scattering amplitude, projected onto the positive
energy states. 
Constructing the quasipotential of the quark-antiquark interaction, 
we have assumed that the effective
interaction is the sum of the usual one-gluon exchange term with the mixture
of long-range vector and scalar linear confining potentials, where
the vector confining potential
contains the Pauli interaction. The quasipotential is then defined by
\footnote{In our notation, where strong annihilation processes are neglected,
 antiparticles are described by usual spinors  taking into account the
 proper quark charges.} 
  \begin{equation}
\label{qpot}
V({\bf p,q};M)=\bar{u}_1(p)\bar{u}_2(-p){\mathcal V}({\bf p}, {\bf
q};M)u_1(q)u_2(-q),
\end{equation}
with
$${\mathcal V}({\bf p},{\bf q};M)
\equiv{\mathcal V}({\bf p}-{\bf q})=\frac{4}{3}\alpha_sD_{ \mu\nu}({\bf
k})\gamma_1^{\mu}\gamma_2^{\nu}
+V^V_{\rm conf}({\bf k})\Gamma_1^{\mu}
\Gamma_{2;\mu}+V^S_{\rm conf}({\bf k}),$$
where $\alpha_s$ is the QCD coupling constant, $D_{\mu\nu}$ is the
gluon propagator in the Coulomb gauge
\begin{equation}
D^{00}({\bf k})=-\frac{4\pi}{{\bf k}^2}, \quad D^{ij}({\bf k})=
-\frac{4\pi}{k^2}\left(\delta^{ij}-\frac{k^ik^j}{{\bf k}^2}\right),
\quad D^{0i}=D^{i0}=0,
\end{equation}
and ${\bf k=p-q}$; $\gamma_{\mu}$ and $u(p)$ are 
the Dirac matrices and spinors
\begin{equation}
\label{spinor}
u^\lambda({p})=\sqrt{\frac{\epsilon(p)+m}{2\epsilon(p)}}
\left(
\begin{array}{c}1\cr {\displaystyle\frac{\bm{\sigma}
      {\bf  p}}{\epsilon(p)+m}}
\end{array}\right)\chi^\lambda,
\end{equation}
with $\epsilon(p)=\sqrt{p^2+m^2}$.
The effective long-range vector vertex is
given by
\begin{equation}
\label{kappa}
\Gamma_{\mu}({\bf k})=\gamma_{\mu}+
\frac{i\kappa}{2m}\sigma_{\mu\nu}k^{\nu},
\end{equation}
where $\kappa$ is the Pauli interaction constant characterizing the
anomalous chromomagnetic moment of quarks. Vector and
scalar confining potentials in the nonrelativistic limit reduce to
\begin{eqnarray}
\label{vlin}
V^V_{\rm conf}(r)&=&(1-\varepsilon)(Ar+B),\nonumber\\ 
V^S_{\rm conf}(r)& =&\varepsilon (Ar+B),
\end{eqnarray}
reproducing 
\begin{equation}
\label{nr}
V_{\rm conf}(r)=V^S_{\rm conf}(r)+V^V_{\rm conf}(r)=Ar+B,
\end{equation}
where $\varepsilon$ is the mixing coefficient. 

All the model parameters have the same values as in our previous
papers \cite{egf,efg}.
The light constituent quark masses $m_u=m_d=0.33$ GeV, $m_s=0.5$ GeV and
the parameters of the linear potential $A=0.18$ GeV$^2$ and $B=-0.3$ GeV
have the usual values of quark models.  The value of the mixing
coefficient of vector and scalar confining potentials $\varepsilon=-1$
has been determined from the consideration of charmonium radiative
decays \cite{efg}. 
Finally, the universal Pauli interaction constant $\kappa=-1$ has been
fixed from the analysis of the fine splitting of heavy quarkonia ${
}^3P_J$- states \cite{efg}. In the literature it is widely discussed
the 't~Hooft-like interaction between quarks induced by instantons \cite{dk}.
This interaction can be partly described by introducing the quark
anomalous chromomagnetic moment having an approximate value
$\kappa=-0.744$ (Diakonov \cite{dk}). This value is of the same
sign and order of magnitude 
as the Pauli constant $\kappa=-1$ in our model. Thus the Pauli term
incorporates at least part of the instanton contribution to the $q\bar q$
interaction.\footnote{As is well-known, the instanton-induced 't~Hooft
  interaction term breaks the axial $U_A(1)$-symmetry, the violation
of  which is needed for  describing the $\eta-\eta'$ mass splitting. We
do not consider this issue here.}

\section{Light meson masses}
\label{sec:lmm}
The quasipotential (\ref{qpot}) can  be used for arbitrary quark
masses.  The substitution 
of the Dirac spinors (\ref{spinor}) into (\ref{qpot}) results in an extremely
nonlocal potential in the configuration space. Clearly, it is very hard to 
deal with such potentials without any additional transformations.
 In oder to simplify the relativistic $q\bar q$ potential, we make the
following replacement in the Dirac spinors:
\begin{equation}
  \label{eq:sub}
  \epsilon_{1,2}(p)=\sqrt{m_{1,2}^2+{\bf p}^2} \to E_{1,2}
\end{equation}
(see the discussion of this point in \cite{egf,lmm}).  This substitution
makes the Fourier transformation of the potential (\ref{qpot}) local.
We also limit our consideration only to the $S$-wave
states, which further simplifies our analysis, since all terms 
proportional to ${\bf L}^2$ vanish as well as the spin-orbit
ones. Thus we neglect the mixing of states with different values of
$L$.  Calculating the potential, we keep only  operators quadratic
in the relative momentum acting on $V_{\rm Coul}$, $V^{V,S}_{\rm conf}$  and
replace ${\bf p}^2\to E_{1,2}^2-m_{1,2}^2$ in higher order operators
in accord  with Eq.~(\ref{eq:sub}) preserving the symmetry under the
$(1\leftrightarrow 2)$ exchange.  

The substitution (\ref{eq:sub})
works well for the confining part of the potential. However, it leads to 
a fictitious singularity $\delta^3({\bf r})$  at the origin arising from the  
one-gluon exchange part ($\Delta V_{\rm
  Coul}(r)$), which is absent in the initial potential.
Note that this singularity is not important if it is treated
perturbatively. Since we are not using the  expansion in $v/c$ and
are solving the quasipotential equation with the 
complete relativistic potential, an additional analysis is
required. Such singular contributions emerge from the following  terms  
\begin{eqnarray}
  \label{eq:st}
 && \frac{{\bf k}^2}{[\epsilon_i(q)(\epsilon_i(q)+m_i)
\epsilon_i(p)(\epsilon_i(p)+m_i)]^{1/2}}V_{\rm Coul}({\bf k}^2) ,\cr
&&\frac{{\bf k}^2}{[\epsilon_1(q)\epsilon_1(p)
\epsilon_2(q)\epsilon_2(p)]^{1/2}}V_{\rm Coul}({\bf k}^2),
\end{eqnarray}
if we simply apply the  replacement (\ref{eq:sub}). However, the Fourier
transforms of expressions (\ref{eq:st}) are less singular at $r\to
0$. To avoid such fictitious singularities we note that if the binding effects 
are taken into account, it is necessary to replace $\epsilon_{1,2}
\to E_{1,2}-\eta_{1,2}V$, where $V$ is the quark interaction potential
and $\eta_{1,2}=m_{2,1}/(m_1+m_2)$. At small
distances  $r\to 0$, the Coulomb singularity in $V$ dominates
and affords the correct asymptotic behaviour. Therefore, we replace
$\epsilon_{1,2} \to E_{1,2}-\eta_{1,2}V_{\rm Coul}$  in  the Fourier
transforms of terms (\ref{eq:st}) (cf. \cite{bs}). We used
the similar regularization of singularities in the analysis of
heavy-light meson spectra \cite{egf}. Finally, we ignore the annihilation
terms in the quark potential since they contribute only in the
isoscalar channels and are suppressed in the $s\bar s$ vector channel
\cite{gi}.

The resulting $q\bar q$ potential then reads
\begin{equation}
  \label{eq:v}
  V(r)= V_{\rm SI}(r)+ V_{\rm SD}(r),
\end{equation}
where the spin-independent potential for $S$-states (${\bf
  L}^2=0$) has the form 
\begin{eqnarray}
  \label{eq:vsi}
  V_{\rm SI}(r)&=&V_{\rm Coul}(r)+V_{\rm conf}(r)+
\frac{(E_1^2-m_1^2+E_2^2-m_2^2)^2}{4(E_1+m_1)(E_2+m_2)}\Biggl\{
\frac1{E_1E_2}V_{\rm Coul}(r)\cr
&& +\frac1{m_1m_2}\Biggl(1+(1+\kappa)\Biggl[(1+\kappa)\frac{(E_1+m_1)(E_2+m_2)}
{E_1E_2}\cr
&&-\left(\frac{E_1+m_1}{E_1}+\frac{E_1+m_2}{E_2}\right)\Biggr]\Biggr)
V^V_{\rm conf}(r)
+\frac1{m_1m_2}V^S_{\rm conf}(r)\Biggr\}\cr
&&+\frac14\left(\frac1{E_1(E_1+m_1)}\Delta
\tilde V^{(1)}_{\rm Coul}(r)+\frac1{E_2(E_2+m_2)}\Delta
\tilde V^{(2)}_{\rm Coul}(r)\right)\cr
&&-\frac14\left[\frac1{m_1(E_1+m_1)}+\frac1{m_2(E_2+m_2)}-(1+\kappa)
\left(\frac1{E_1m_1}+\frac1{E_2m_2}\right)\right]\Delta V^V_{\rm
conf}(r)\cr
&&+\frac{(E_1^2-m_1^2+E_2^2-m_2^2)}{8m_1m_2(E_1+m_1)(E_2+m_2)} 
\Delta V^S_{\rm conf}(r), 
\end{eqnarray}
and the spin-dependent potential is given by
\begin{eqnarray}
  \label{eq:vsd}
   V_{\rm SD}(r)&=&\frac2{3E_1E_2}\Biggl[\Delta \bar V_{\rm Coul}(r)
+\left(\frac{E_1-m_1}{2m_1}-(1+\kappa)\frac{E_1+m_1}{2m_1}\right)\cr
&&\qquad\quad\times
\left(\frac{E_2-m_2}{2m_2}-(1+\kappa)\frac{E_2+m_2}{2m_2}\right)
\Delta V^V_{\rm conf}(r)\Biggr]{\bf S}_1{\bf S}_2,
\end{eqnarray}
with
\begin{eqnarray}
  \label{eq:tv}
V_{\rm Coul}(r)&=&-\frac43\frac{\alpha_s}{r},\cr
\tilde V^{(i)}_{\rm Coul}(r)&=&V_{\rm Coul}(r)\frac1{\displaystyle\left(1+
\eta_i\frac43\frac{\alpha_s}{E_i}\frac1{r}\right)\left(1+
\eta_i\frac43\frac{\alpha_s}{E_i+m_i}\frac1{r}\right)},\qquad (i=1,2),\cr
  \bar V_{\rm Coul}(r)&=&V_{\rm Coul}(r)\frac1{\displaystyle\left(1+
\eta_1\frac43\frac{\alpha_s}{E_1}\frac1{r}\right)\left(1+
\eta_2\frac43\frac{\alpha_s}{E_2}\frac1{r}\right)}, 
\qquad \eta_{1,2}=\frac{m_{2,1}}{m_1+m_2}.
\end{eqnarray}
Here we put  $\alpha_s\equiv\alpha_s(\mu_{12}^2)$ with $\mu_{12}=2m_1
m_2/(m_1+m_2)$. We adopt for $\alpha_s(\mu^2)$ the
simplest model with freezing \cite{bvb}, namely
\begin{equation}
  \label{eq:alpha}
  \alpha_s(\mu^2)=\frac{4\pi}{\displaystyle\beta_0
\ln\frac{\mu^2+M_B^2}{\Lambda^2}}, \qquad \beta_0=11-\frac23n_f,
\end{equation}
where the background mass is $M_B=2.24\sqrt{A}=0.95$~GeV \cite{bvb}, and
$\Lambda=413$~MeV was fixed from fitting the $\rho$
mass.~\footnote{The definition (\ref{eq:alpha}) of $\alpha_s$ can be
  smoothly matched with the $\alpha_s$ used for heavy quarkonia
  \cite{efg} at the scale about $m_c$.} We put the
number of flavours $n_f=2$ for $\pi$, 
$\rho$, $K$, $K^*$ and $n_f=3$ for $\phi$. As a result we obtain
$\alpha_s(\mu_{ud}^2)=0.730$, $\alpha_s(\mu_{us}^2)=0.711$ and
$\alpha_s(\mu_{ss}^2)=0.731$.

\begin{table}
  \caption{Masses of light $S$-wave mesons (in MeV)}
  \label{tab:mass}
\begin{ruledtabular}
\begin{tabular}{ccccccc}
Meson& State & 
\multicolumn{4}{l}{\underline{\hspace{3.5cm}Theory\hspace{3.5cm}}}
\hspace{-3.6cm}
& Experiment \\
 & $n^{2S+1}L_J$&this work& \cite{gi}& \cite{mr}& \cite{k}&
 PDG \cite{pdg}\\ 
\hline
$\pi$&  $1^1S_0$& 154 &150& 138& 140 &139.57\\
$\rho$ & $1^3S_1$ & 776$^\dag$ & 770& 742& 785 & 775.8(5)\\
$\pi'$&  $2^1S_0$& 1292 &1300& &1331  &1300(100)\\
$\rho'$ & $2^3S_1$ & 1486&1450& &1420 & 1465(25)\\
$\pi''$&$3^1S_0$& 1788 &1880& &1826& 1812(14)\\
$\rho''$ & $3^3S_1$ & 1921 & 2000 & & 1472& \\
$K$ & $1^1S_0$ & 482&470& 497& 506& 493.677(16)\\
$K^*$&$1^3S_1$ & 897&900& 936& 890& 891.66(26)\\
$K'$ & $2^1S_0$ & 1538&1450& &1470& \\
${K^*}'$&$2^3S_1$ & 1675&1580& &1550& 1717(27)\\
$K''$ & $3^1S_0$ & 2065 & 2020& &1965& \\
${K^*}''$&$3^3S_1$ &2156 & 2110& & 1588& \\
$\phi$& $1^3S_1$& 1038&1020& 1072&990 & 1019.46(2)\\
$\phi'$& $2^3S_1$& 1698&1690& &1472 & 1680(20)
  \end{tabular}
\end{ruledtabular}
\flushleft${}^\dag$ fitted value
\end{table}

The quasipotential equation (\ref{quas}) is solved numerically for the
complete relativistic potential (\ref{eq:v}) which depends on the
meson mass in a complicated highly nonlinear way.  The obtained meson
masses are presented in Table~\ref{tab:mass} in comparison with
experimental data \cite{pdg} and other theoretical results \cite{gi,mr,k}.
This comparison exhibits a reasonably good overall agreement of our
predictions with experimental mass values.
Our results are also consistent with mass formulas derived using
the finite-energy sum rules in QCD \cite{kp} and with predictions of
lattice QCD \cite{ak}.   
We consider such agreement to be quite
successful, since in evaluating the meson masses we had at our disposal
only one adjustable parameter  $\Lambda$, which was fixed from fitting
the $\rho$ meson mass. All other parameters are kept the same as in
our previous papers \cite{egf,efg}. The obtained wave functions of the
light mesons are used for the calculation of their decay constants and
electromagnetic form factors in the following sections.

\section{Decay constants}
\label{sec:dc}
\begin{figure}
  \centering
  \includegraphics[width=16cm]{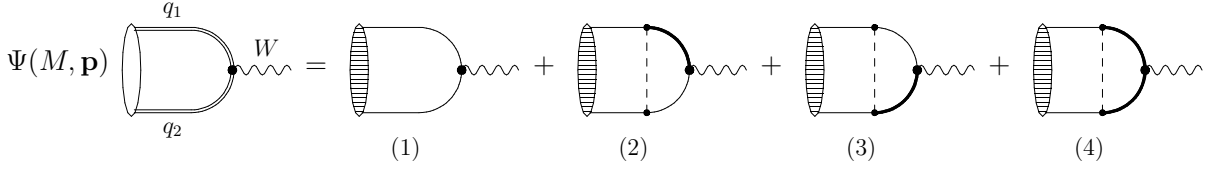}
  \caption{Weak annihilation diagram of the light meson. Solid and bold
    lines denote the  positive- and negative-energy part of the quark propagator,
respectively. Dashed lines represent the interaction operator
${\mathcal V}$.} 
  \label{fig:diag}
\end{figure}

The decay constants $f_P$ and $f_V$ of the pseudoscalar ($P$) and
vector ($V$) mesons parameterize the matrix elements of the weak
current $J^W_\mu=\bar q_1{\cal J}^W_\mu q_2=\bar q_1\gamma_\mu(1-\gamma_5)q_2$
between the corresponding 
meson and the vacuum. They are defined by  
\begin{eqnarray}
  \label{eq:dc}
  \left<0|\bar q_1 \gamma^\mu\gamma_5 q_2|P({\bf K})\right>&=& i f_P
  K^\mu,\\ 
\left<0|\bar q_1 \gamma^\mu q_2|V({\bf K},\varepsilon)\right>&=& f_V
  M_V \varepsilon^\mu,
\end{eqnarray}
where ${\bf K}$ is the meson momentum, $\varepsilon^\mu$ and $M_V$ are
the polarization vector and mass of the vector meson. This matrix
element can be expressed through the two-particle Bethe-Salpeter wave
function in the quark loop integral (see Fig.~\ref{fig:diag})
\begin{equation}
  \label{eq:bs}
  \left<0| J^W_\mu |M({\bf K})\right>=\int\frac{d^4
    p}{(2\pi)^4}
{\rm Tr}\left\{\gamma_\mu(1-\gamma_5)\Psi(M,p)\right\},
\end{equation}
where the trace is taken over spin indices. Integration 
over $p^0$ in Eq.~(\ref{eq:bs}) allows one to pass to the single-time
wave function in the meson rest frame
\begin{equation}
  \label{eq:stw}
  \Psi(M,{\bf p})=\int\frac{dp^0}{2\pi}\Psi(M,p).
\end{equation}
This wave function contains both positive- and negative-energy
quark states. Since in the quasipotential approach we use the single-time wave
function $\Psi_{M\, {\bf K}}({\bf p})$  projected onto the positive-energy states  it
is necessary to 
include additional terms which account for the contributions
of negative-energy intermediate states. The weak annihilation amplitude
(\ref{eq:bs}) is schematically presented in the left hand side of
Fig.~\ref{fig:diag}. The first diagram on the right hand side
corresponds to the simple replacing of the single-time wave function
(\ref{eq:stw})  $\Psi(M,{\bf p})$ by the quasipotential one
$\Psi_{M\, {\bf K}}({\bf p})$.\footnote{The contributions with the exchange by the
  effective interaction potential ${\mathcal V}$ which contain only
  positive-energy intermediate states are automatically accounted for
  by the wave function itself.}  The second and third diagrams account for
negative-energy contributions to the first and second quark propagators,
respectively. The last diagram corresponds to negative-energy
contributions from both quark propagators. 

 Thus in the quasipotential approach
this decay amplitude has the form
\begin{eqnarray}
  \label{eq:qpd}
\left<0|J^W_\mu |M({\bf K})\right>&=&\sqrt{2M}\Biggl\{\int\frac{d^3
  p}{(2\pi)^3} \bar u_1(p_1){\cal J}^W_\mu u_2(p_2)
 \Psi_{M\, {\bf K}}({\bf
  p})+\Biggl[\int\frac{d^3 p d^3 p'}{(2\pi)^6}\bar u_1(p_1)\Gamma_1\cr 
&&\!\!\!\!\times
\frac{\Lambda^{(-)}_1(p_1')\gamma^0{\cal J}^W_\mu\Lambda^{(+)}_2(p_2')\gamma^0}
{M+\epsilon_1(p')-\epsilon_2(p')}\Gamma_2 u_2(p_2)\tilde V(p-p')
\Psi_{M\, {\bf K}}({\bf
  p})
+(1\leftrightarrow 2)\Biggr]\cr
&&\!\!\!\!+\int\frac{d^3 p d^3 p'}{(2\pi)^6}\bar u_1(p_1)\Gamma_1
\frac{\Lambda^{(-)}_1(p_1')\gamma^0{\cal J}^W_\mu\Lambda^{(-)}_2(p_2')\gamma^0}
{M+\epsilon_1(p')+\epsilon_2(p')}\Gamma_2 u_2(p_2)\tilde V(p-p')
\Psi_{M\, {\bf K}}({\bf
  p})\Biggr\},\cr&&\!\!\!
\end{eqnarray}
where ${\bf p}_{1,2}^{(')}={\bf K}/2\pm {\bf p}^{(')}$; matrices
$\Gamma_{1,2}$ denote the Dirac structure of the interaction 
potential (\ref{qpot}) for the first and second quark, respectively, and thus
$\Gamma_1\Gamma_2\tilde V(p-p')={\mathcal V}({\bf p}-{\bf p}')$. The
factor $\sqrt{2M}$ follows
from the normalization of the quasipotential wave function. The
positive- and negative-energy projectors have standard definition
\[
\Lambda^{(\pm)}(p)={\epsilon(p)\pm\bigl( m\gamma ^0+\gamma^0(
\bm{\gamma}{\bf p})\bigr) \over 2\epsilon (p)}.
\]  
The quasipotential wave function  in the rest frame
of the decaying meson $\Psi_M({\bf p})\equiv\Psi_{M\, {\bf 0}}({\bf p})$
can be expressed through a product of radial $\Phi_M(p)$,
spin $\chi_{ss'}$ and colour $\phi_{q_1q_2}$ wave functions
\begin{equation}
  \label{eq:wff}
  \Psi_M({\bf p})=\Phi_M(p)\chi_{ss'}\phi_{q_1q_2}.
\end{equation}

Now the decay constants can be presented in the following form
\begin{equation}
  \label{eq:fpe}
  f_{P,V}=f_{P,V}^{(1)}+f_{P,V}^{(2+3)}+f_{P,V}^{(4)},
\end{equation}
where the terms on the right hand side originate from the corresponding
diagrams in Fig.~\ref{fig:diag} and parameterize respective terms in
Eq.~(\ref{eq:qpd}). In the literature \cite{gi,g,vd,efgdc} usually only
the first term is taken 
into account, since it provides the nonrelativistic limit, while
other terms give only relativistic corrections and thus vanish in
this limit. Such approximation can be justified for  mesons
containing heavy quarks. However,  as it will be shown below, for
light mesons other terms become equally important and their account is
crucial for getting the results in agreement with experimental data. 

The matrix element (\ref{eq:qpd}) and thus the decay constants can be
calculated in an arbitrary  
frame and from any component of the weak current. Such calculation can
be most easily performed in the rest frame of the decaying meson from the
zero component of the current. The same results will be obtained from the
vector component, however, this calculation is more cumbersome since
here the rest frame cannot be used and, thus, it is important to take
into account the relativistic transformation of the meson wave function
from the rest frame to the moving one with the momentum ${\bf K}$ (see
Eq.~(\ref{wig}) below). It
is also possible to perform calculations in the explicitly covariant
way using methods proposed in \cite{efgms}.   

The resulting expressions for decay constants are given by
\begin{eqnarray}
  \label{eq:fpv1}
  f^{(1)}_{P,V}&=&\sqrt{\frac{12}{M}}\int \frac{d^3
  p}{(2\pi)^3}\left(\frac{\epsilon_1(p)+m_1}{2\epsilon_1(p)}\right)^{1/2}
  \left(\frac{\epsilon_2(p)+m_2}{2\epsilon_2(p)}\right)^{1/2}
 \cr
&&\times \left\{ 1
  +\lambda_{P,V}\,\frac{{\bf p}^2}{[\epsilon_1(p)+m_1][\epsilon_2(p)+m_2]}\right\}
  \Phi_{P,V}(p),
\end{eqnarray}
\begin{eqnarray}
  \label{eq:fpv3}
  f^{(2+3)}_{P,V}&=&\sqrt{\frac{12}{M}}\int \frac{d^3
  p}{(2\pi)^3}\left(\frac{\epsilon_1(p)+m_1}{2\epsilon_1(p)}\right)^{1/2}
  \left(\frac{\epsilon_2(p)+m_2}{2\epsilon_2(p)}\right)^{1/2}\Biggl[
\frac{M-\epsilon_1(p)-\epsilon_2(p)}{M+\epsilon_1(p)-\epsilon_2(p)}\cr
&&\times
\frac{{\bf p}^2}{\epsilon_1(p)[\epsilon_1(p)+m_1]}
  \left\{1+\lambda_{P,V}\frac{\epsilon_1(p)+m_1}{\epsilon_2(p)+m_2}\right\}
+(1\leftrightarrow 2)\Biggr]
  \Phi_{P,V}(p),
\end{eqnarray}
\begin{eqnarray}
  \label{eq:fpv4}
  f^{(4)}_{P,V}&=&\sqrt{\frac{12}{M}}\int \frac{d^3
  p}{(2\pi)^3}\left(\frac{\epsilon_1(p)+m_1}{2\epsilon_1(p)}\right)^{1/2}
  \left(\frac{\epsilon_2(p)+m_2}{2\epsilon_2(p)}\right)^{1/2}
\frac{M-\epsilon_1(p)-\epsilon_2(p)}{M+\epsilon_1(p)+\epsilon_2(p)}\cr
&&\times
  \left\{-\lambda_{P,V}-\frac{{\bf p}^2}{[\epsilon_1(p)+m_1][\epsilon_2(p)+m_2]}\right\} \cr
&&\times
\left[\frac{(1-\varepsilon)m_1^2m_2^2}{\epsilon_1^2(p)\epsilon_2^2(p)}+
\frac{{\bf p}^2}{[\epsilon_1(p)+m_1][\epsilon_2(p)+m_2]}\right]
  \Phi_{P,V}(p),
\end{eqnarray}
with $\lambda_P=-1$ and $\lambda_V=1/3$. Here $\varepsilon$ is the
mixing coefficient of scalar and vector confining potentials
(\ref{vlin}) and the long-range anomalous chromomagnetic quark moment
$\kappa$ (\ref{kappa}) is put equal to $-1$. Note that  $f_P^{(2+3)}$
vanishes for pseudoscalar mesons with equal quark masses, such as the pion.
The positive-energy contribution (\ref{eq:fpv1}) reproduces the previously known
expressions for the decay constants \cite{gi,g}. The negative-energy
contributions (\ref{eq:fpv3}) and (\ref{eq:fpv4}) are new and play a
significant role for light mesons (see below).

In the nonrelativistic limit $p^2/m^2\to 0$ the expression (\ref{eq:fpv1}) for
decay constants gives the well-known formula
\begin{equation}
\label{eq:fnr}
f_{P,V}^{\rm NR}=
\sqrt{\frac{12}{M_{P,V}}}\left|\Psi_{P,V}(0)\right|,
\end{equation}
where $\Psi_{P,V}(0)$ is the meson wave function at the origin
$r=0$. All other contributions vanish in the nonrelativistic limit.

\begin{table}
  \caption{Different contributions to the pseudoscalar and vector
    decay constants of light mesons (in  MeV). The notations are taken
  according to Eqs.~(\ref{eq:fpe}) and (\ref{eq:fnr}).} 
  \label{tab:dc}
\begin{ruledtabular}
\begin{tabular}{cccccc}
Constant& $f_M^{\rm NR}$&$f_M^{(1)}$& $f_M^{(2+3)}+f_M^{(4)}$
& $(f_M^{(2+3)}+f_M^{(4)})/f^{(1)}_M $&$f_M$ \\
\hline
$f_\pi$ & 1290 &515 & $-391$ &$-76\%$ &124  \\
$f_K$ & 783 & 353 & $-198$& $-56\%$ & 155 \\
$f_\rho$ & 490 & 402 & $-183$&$-46\%$ & 219 \\
$f_{K^*}$ & 508 & 410 & $-174$&$-42\%$ & 236\\
$f_\phi$ & 511 & 415 &$-170$&$-41\%$  &245 
\end{tabular}
\end{ruledtabular}

\end{table}

\begin{table}
  \caption{Pseudoscalar and vector decay constants of light mesons (in
    MeV).}
  \label{tab:dce}
\begin{ruledtabular}
\begin{tabular}{ccccccccc}
Constant&this work&\cite{gi} & \cite{mr,mt} &\cite{k} &\cite{hjd}
&Lattice \cite{ak} &Lattice \cite{milc}& Experiment \cite{pdg}\\
\hline
$f_\pi$ & 124& 180 & 131&219&138&$126.6\pm6.4$ &$129.5\pm3.6$  &$130.7\pm0.1\pm0.36$\\
$f_K$ & 155& 232& 155& 238&160&$152.0\pm6.1$&$156.6\pm3.7$ & $159.8\pm1.4\pm0.44$\\
$f_\rho$ & 219&220&207& &238&$239.4\pm7.3$& & $220\pm2^*$\\
$f_{K^*}$ & 236& 267& 241& &241&$255.5\pm6.5$& & $230\pm8^\dag$\\
$f_\phi$ &245&336&259&&&$270.8\pm6.5$& & $229\pm3^\ddag$  
\end{tabular}
\end{ruledtabular}
\begin{flushleft}
${}^*$ derived from the experimental value for $\Gamma_{\rho^0\to
  e^+e^-}$.\\
${}^\dag$ derived from the experimental value for the ratio
$\Gamma_{\tau\to K^*\nu_\tau}/\Gamma_{\tau\to\rho\nu_\tau}$ and the
$f_\rho$ value.\\
${}^\ddag$ derived from the experimental value for $\Gamma_{\phi\to
  e^+e^-}$. 
\end{flushleft}

\end{table}

In Table~\ref{tab:dc} we present our predictions for the
light meson decay constants calculated using the meson wave functions
which were obtained as the numerical solutions of the quasipotential equation in
Sec.~\ref{sec:lmm}. The nonrelativistic values $f_M^{\rm NR}$
(\ref{eq:fnr}) as well as the values of different contributions in
Fig.~\ref{fig:diag} $f_M^{(1,2,3,4)}$ (\ref{eq:fpv1})--(\ref{eq:fpv4})
and the full relativistic results $f_M$ (\ref{eq:fpe})  are given. In
Table~\ref{tab:dce} we compare our results  for the decay constants $f_M$
with predictions of other approaches \cite{gi,mr,mt,k,hjd}, recent values from
two- \cite{ak} and three-flavour lattice QCD \cite{milc}  and
available experimental data \cite{pdg}. It is clearly seen that the
nonrelativistic predictions are significantly overestimating all
decay constants, especially for the pion (almost by a factor of 10). The
account of the part of relativistic corrections by keeping in
Eq.~(\ref{eq:fpe}) only the first term $f_M^{(1)}$
(\ref{eq:fpv1}), which is usually used for semirelativistic
calculations,  does not dramatically improve the situation. The
disagreement is still large.  This is connected with the anomalously
small masses of light pseudoscalar mesons exhibiting their chiral
nature. In the semirelativistic quark model \cite{gi,g} the pseudoscalar
meson mass is replaced by the so-called mock mass $\tilde M_P$, which is equal to
the mean total energy of free quarks in a meson, and with our wave
functions: $\tilde
M_\pi=2\langle\epsilon_q(p)\rangle\approx 1070$~MeV ($\sim 8 M_\pi$)
and $\tilde 
M_K=\langle\epsilon_q(p)\rangle+\langle\epsilon_s(p)\rangle\approx
1232$~MeV ($\sim 2.5 M_K$). Such replacement gives $f_P^{(1)}$ values
which are still $\approx 1.4$ times larger than experimental ones
(cf. \cite{gi}). As we see from Table~\ref{tab:dc}, in the
quasipotential approach it is not justified to neglect contributions of
the negative energy intermediate states for light meson decay
constants. Indeed, the values of $f_M^{(2+3)}+f_M^{(4)}$ are large and
negative (reaching $-76\%$ of $f_\pi^{(1)}$ for the pion) 
thus compensating the overestimation of decay constants by
the positive-energy contribution $f_M^{(1)}$. This is the consequence
of the smallness of the 
light pseudoscalar meson masses compared to the energies of their
constituents. The negative-energy contributions (\ref{eq:fpv3}),
(\ref{eq:fpv4}) are proportional to the ratio of the meson binding
energy $M-\epsilon_1(p)-\epsilon_2(p)$ to its mass.  For mesons with
heavy quarks this factor leads to the suppression of negative-energy
contributions since the binding energies are small on the heavy
meson mass scale. This results in the dominance of the positive-energy
term $f_M^{(1)}$ since the negative-energy terms give only $1/m_Q$
contributions ($m_Q$ is the heavy quark mass).\footnote{For the
  heavy-heavy $B_c$ meson ($c\bar b$) these negative-energy
  corrections will be of order $v^4/c^4$ and thus very small. The
  influence of the  
  negative-energy contributions $f_M^{(2+3,4)}$ on the decay constants
  of heavy-light $B$ and $D$ mesons will be considered elsewhere.}  
On the other hand, for light mesons,
especially for the pion and kaon, the binding energies are large on the
light meson mass scale and, thus, such factor gives no
suppression. Taking the complete relativistic expression for decay
constants $f_M$ (\ref{eq:fpe}) brings theoretical predictions in good
agreement with available experimental data. 

The comparison of our values of the decay constants with other predictions in
Table~\ref{tab:dce} indicate that they are competitive even with the results of
more sophisticated approaches \cite{mt,k} which are based on the
Dyson-Schwinger and Bethe-Salpeter equations. On the other hand our
model is more selfconsistent than some other approaches \cite{hjd,kt,gi,g,vd}. 
We calculate the meson wave functions by solving the
quasipotential equation in contrast to the models based on the
relativistic Hamilton dynamics \cite{hjd,kt} where various ad hoc
wave function parameterizations are employed. We also do not need to
introduce the mock meson mass \cite{gi,g,vd} and to substitute it for the
light meson mass as it was discussed above.

\section{ Electromagnetic form factors}
\label{em}

The elastic matrix element of the
electromagnetic current $J_\mu$ between the initial  and final
pseudoscalar meson states is parameterized by the form factor $F_P(Q^2)$ 
\begin{equation}
  \label{eq:sff}
  \langle M(P_F)\vert J_\mu \vert M(P_I)\rangle=F_P(Q^2)(P_I+P_F)_\mu,
\end{equation}
where $Q^2=-(P_F-P_I)^2$.

\begin{figure}
  \centering
  \includegraphics{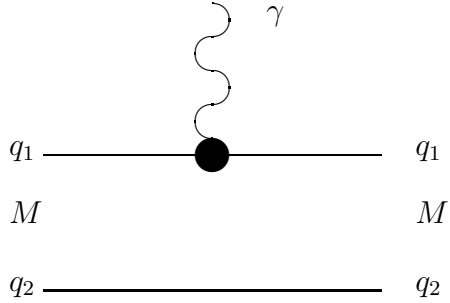}
  \caption{Lowest order vertex function $\Gamma^{(1)}$
corresponding to Eq.~(\ref{gam1}). Photon interaction with one quark is shown.}
  \label{fig:1}
\end{figure}

\begin{figure}
  \centering
  \includegraphics{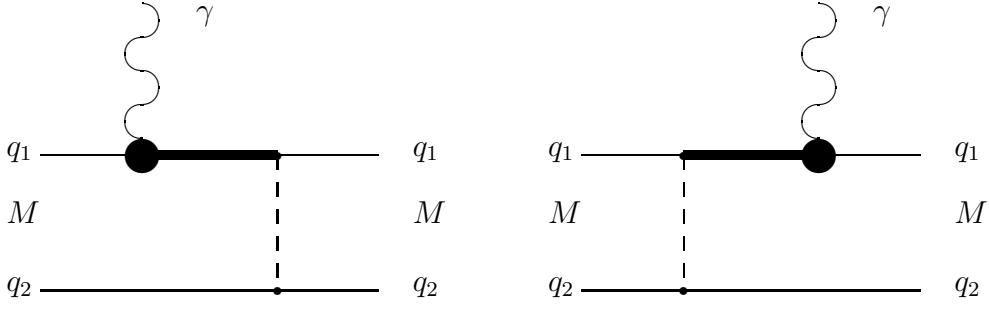}
  \caption{ Vertex function $\Gamma^{(2)}$
corresponding to Eq.~(\ref{gam2}). Dashed lines represent the interaction 
operator ${\mathcal V}$. Bold lines denote the  
negative-energy part of the quark propagator. As on Fig.~\ref{fig:1},
photon interaction with one quark is shown.}
  \label{fig:2}
\end{figure}

In the quasipotential approach such  matrix element has the form \cite{f}
\begin{equation}
\label{mxet}
\langle M(P_F) \vert J_\mu  \vert M(P_I)\rangle
=\int \frac{d^3p\, d^3q}{(2\pi )^6} \bar \Psi_{M\, {\bf P}_F}({\bf
p})\Gamma _\mu ({\bf p},{\bf q})\Psi_{M\, {\bf P}_I}({\bf q}),
\end{equation}
where $\Gamma _\mu ({\bf p},{\bf
q})$ is the two-particle vertex function and  $\Psi_{M}$ are the
meson wave functions projected onto the positive energy states of
quarks and boosted to the moving reference frame.
The contributions to $\Gamma$ come from Figs.~\ref{fig:1} and \ref{fig:2}.
The term $\Gamma^{(2)}$ includes contributions from the
negative-energy quark states. Note that the form
of the relativistic corrections resulting from the vertex function
$\Gamma^{(2)}$ explicitly depends on the Lorentz structure of the
$q\bar q$-interaction.  Thus the vertex function is given by
\begin{equation}
  \label{eq:gam}
  \Gamma_\mu({\bf p},{\bf q})=\Gamma_\mu^{(1)}({\bf p},{\bf q})+
  \Gamma_\mu^{(2)}({\bf p},{\bf q})+ \cdots ,
\end{equation}
where
\begin{equation}\label{gam1}
\Gamma_\mu ^{(1)}({\bf p},{\bf q})=e_1\bar
u_1(p_1)\gamma_\mu  
u_1(q_1)(2\pi)^3\delta({\bf p}_2-{\bf q}_2) +(1\leftrightarrow 2),
\end{equation}
and
\begin{eqnarray}\label{gam2}
\Gamma_\mu^{(2)}({\bf p},{\bf q})&=&e_1\bar u_1(p_1)\bar
u_2(p_2) \biggl\{{\mathcal V}({\bf p}_2-{\bf q}_2)
\frac{\Lambda_1^{(-)}(k_1')}{
\epsilon_1(k_1')+ \epsilon_1(q_1)}\gamma_1^0\gamma_{1\mu}
\nonumber\\
& & + \gamma_{1\mu}
\frac{\Lambda_1^{(-)}({k}_1)}{ \epsilon_1(k_1)+\epsilon_1(p_1)}
\gamma_1^0{\mathcal V}({\bf p}_2-{\bf q}_2)
\biggr\}u_1(q_1) u_2(q_2) +(1\leftrightarrow 2). 
\end{eqnarray}
Here $e_{1,2}$ are the quark charges, ${\bf k}_1={\bf p}_1-{\bf\Delta};
\quad {\bf k}_1'={\bf 
q}_1+{\bf\Delta};\quad {\bf\Delta}={\bf P}_F-{\bf P}_I$; 
\[
\Lambda^{(-)}(p)={\epsilon(p)-\bigl( m\gamma ^0+\gamma^0(
\bm{\gamma}{\bf p})\bigr) \over 2\epsilon (p)}, \qquad \epsilon(p)=
\sqrt{p^2+m^2},
\]
and  
\begin{eqnarray*} 
p_{1,2}&=&\epsilon_{1,2}(p)\frac{P_{F}}{M}
\pm\sum_{i=1}^3 n^{(i)}(P_{F})p^i,\\
q_{1,2}&=&\epsilon_{1,2}(q)\frac{P_I}{M} \pm \sum_{i=1}^3 n^{(i)}
(P_I)q^i,\end{eqnarray*}
where $n^{(i)}$ are three four-vectors given by
$$ n^{(i)\mu}(p)=\left\{ \frac{p^i}{M},\ \delta_{ij}+
\frac{p^ip^j}{M(E+M)}\right\}, \quad E=\sqrt{{\bf p}^2+M^2},
\qquad i,j=1,2,3,$$
$P_I=(E_I,{\bf P}_I)$ and $P_F=(E_F,{\bf P}_F)$ are four-momenta of
the initial and final mesons.

It is important to note that the wave functions entering the current
matrix element (\ref{mxet}) cannot  be both in the rest frame.
In the initial  meson rest frame, the final meson is moving
with the recoil momentum ${\bf \Delta}$. The wave function
of the moving meson $\Psi_{M\,{\bf\Delta}}$ is connected
with the wave function in the rest frame
$\Psi_{M\,{\bf 0}}\equiv \Psi_{M}$ by the transformation \cite{f}
\begin{equation}
\label{wig}
\Psi_{M\,{\bf\Delta}}({\bf p})
=D_1^{1/2}(R_{L_{\bf\Delta}}^W)D_2^{1/2}(R_{L_{\bf\Delta}}^W)
\Psi_{M\,{\bf 0}}({\bf p}),
\end{equation}
where $R^W$ is the Wigner rotation, $L_{\bf\Delta}$ is the Lorentz boost
from the rest frame to a moving one, and the rotation matrix
$D^{1/2}(R)$ in the spinor representation  is given by 
\begin{equation}\label{d12}
{1 \ \ \,0\choose 0 \ \ \,1}D^{1/2}_{1,2}(R^W_{L_{\bf\Delta}})=
S^{-1}({\bf p}_{1,2})S({\bf\Delta})S({\bf p}),
\end{equation}
where
$$
S({\bf p})=\sqrt{\frac{\epsilon(p)+m}{2m}}\left(1+\frac{\bm{\alpha}
{\bf p}} {\epsilon(p)+m}\right)
$$
is the usual Lorentz transformation matrix of the Dirac spinor.

To calculate the matrix element (\ref{eq:sff}) of the electromagnetic
current between the pseudoscalar meson states we substitute the vertex
functions $\Gamma^{(1)}$ (\ref{gam1}) and $\Gamma^{(2)}$ (\ref{gam2})
in Eq.~(\ref{mxet}) and take into account the wave function
transformation (\ref{wig}). Then we use the $\delta$ function in
$\Gamma^{(1)}$ to perform one of the integrations in the matrix
element (\ref{mxet}). For the contribution of $\Gamma^{(2)}$ we use
instead the quasipotential equation to replace the integral of the product
of the interaction potential and the bound state wave function by the
product of the corresponding binding energy and the wave function. To
simplify the calculation we explicitly use the value $\kappa=-1$ for
the long-range anomalous chromomagnetic quark moment
(\ref{kappa}). However, as previously we keep the dependence on the
mixing parameter $\varepsilon$ of the vector and scalar confining
potentials (\ref{vlin}).  As a
result we get the following expression for the electromagnetic form
factor of the pseudoscalar meson:

\begin{equation}
  \label{eq:eff}
  F_P({Q}^2)=F_P^{(1)}({Q}^2)+\varepsilon F_P^{(2)S}({Q}^2)
+(1-\varepsilon)F_P^{(2)V}({Q}^2),
\end{equation}
\begin{eqnarray}\label{eq:f1}
F_P^{(1)}({Q}^2)&=&\frac{2\sqrt{EM}}{E+M}\Biggl\{e_1
  \int \frac{d^3p}{(2\pi )^3} \bar\Psi_{M}
\left({\bf p}+
\frac{2\epsilon_{2}(p)}{E+M}{\bf \Delta } \right)
\sqrt{\frac{\epsilon_1(p)+m_1}{\epsilon_1(p+\Delta)+m_1}}
\Biggl[\frac{\epsilon_1(p+\Delta)+\epsilon_1(p)}
{2\sqrt{\epsilon_1(p+\Delta)\epsilon_1(p)}}\cr
&&+\frac{\bf p \Delta}{2\sqrt{\epsilon_1(p+\Delta)\epsilon_1(p)}
(\epsilon_1(p)+m_1)} -\frac{\epsilon_1(p+\Delta)-\epsilon_1(p)}
{2\sqrt{\epsilon_1(p+\Delta)\epsilon_1(p)}} 
\frac{{\bf p}_T^2}{\epsilon_1(p)+m_1}\cr
&&\times
\left(\frac1{\epsilon_1(p)+m_1}+\frac1{\epsilon_2(p)+m_2}\right)
\Biggr]\Psi_{M}({\bf
  p})+(1\leftrightarrow 2)\Biggr\},
\end{eqnarray}

\begin{eqnarray}\label{eq:f2s}
F_P^{(2)S}({Q}^2)&=&\frac{2\sqrt{EM}}{E+M}\Biggl\{e_1
  \int \frac{d^3p}{(2\pi )^3} \bar\Psi_{M}
\left({\bf p}+
\frac{2\epsilon_{2}(p)}{E+M}{\bf \Delta } \right)
\sqrt{\frac{\epsilon_1(p)+m_1}{\epsilon_1(p+\Delta)+m_1}}
\frac{\epsilon_1(p+\Delta)+m_1}
{2\epsilon_1(p+\Delta)}\cr
&&\times
\Biggl[\frac{\epsilon_1(p+\Delta)-\epsilon_1(p)+2m_1}
{2\sqrt{\epsilon_1(p+\Delta)\epsilon_1(p)}}
-\frac{\bf p \Delta}{2\sqrt{\epsilon_1(p+\Delta)\epsilon_1(p)}
(\epsilon_1(p)+m_1)}\cr
&& -\frac{\epsilon_1(p+\Delta)+m_1}
{2\sqrt{\epsilon_1(p+\Delta)\epsilon_1(p)}} \frac{{\bf
    p}_T^2}{\epsilon_1(p)+m_1}
\left(\frac1{\epsilon_1(p)+m_1}+\frac1{\epsilon_2(p)+m_2}\right)\Biggr]\cr
&&\times
 \frac{\epsilon_1(p+\Delta)-\epsilon_1(p)}{\epsilon_1(p+\Delta)
[\epsilon_1(p+\Delta)+\epsilon_1(p)]}[M-\epsilon_1(p)-\epsilon_2(p)]
\Psi_{M}({\bf
  p})+(1\leftrightarrow 2)\Biggr\},
\end{eqnarray}

\begin{eqnarray}\label{eq:f2v}
F_P^{(2)V}(Q^2)&=&\frac{2\sqrt{EM}}{E+M}\Biggl\{e_1
  \int \frac{d^3p}{(2\pi )^3} \bar\Psi_{M}
\left({\bf p}+
\frac{2\epsilon_{2}(p)}{E+M}{\bf \Delta } \right)
\sqrt{\frac{\epsilon_1(p)+m_1}{\epsilon_1(p+\Delta)+m_1}}
\frac{\epsilon_1(p+\Delta)+m_1}
{2\epsilon_1(p+\Delta)}\cr
&&\times
\Biggl[\frac{\epsilon_1(p)-m_1}
{2\sqrt{\epsilon_1(p+\Delta)\epsilon_1(p)}}
+\frac{\bf p \Delta}{2\sqrt{\epsilon_1(p+\Delta)\epsilon_1(p)}
(\epsilon_1(p)+m_1)}\cr
&& +\frac{\epsilon_1(p+\Delta)+m_1}
{2\sqrt{\epsilon_1(p+\Delta)\epsilon_1(p)}} \frac{{\bf
    p}_T^2}{\epsilon_1(p)+m_1}
\left(\frac1{\epsilon_1(p)+m_1}+\frac1{\epsilon_2(p)+m_2}\right)\Biggr]\cr
&&\times
 \frac{\epsilon_1(p+\Delta)-\epsilon_1(p)}{\epsilon_1(p+\Delta)
[\epsilon_1(p+\Delta)+\epsilon_1(p)]}[M-\epsilon_1(p)-\epsilon_2(p)]
\Psi_{M}({\bf
  p})+(1\leftrightarrow 2)\Biggr\},
\end{eqnarray}
where $F_P^{(2)S(V)}$ are contributions from scalar (vector) confining
potentials and ${\bf p}_T={\bf p}^2-({\bf p\Delta})^2/{\bf \Delta}^2$,
$E=\sqrt{M^2+\bf{\Delta}^2}$. As previously, we put $\varepsilon=-1$
for further numerical 
calculations. It is important to note that the above expressions for
the electromagnetic form factor of the positively-charged pseudoscalar meson
exactly satisfy the normalization condition 
\begin{equation}
  \label{eq:nc}
 F_P(0)=1
\end{equation}
following from the electric charge conservation.

\begin{figure}
  \centering
\includegraphics[height=12cm,angle=-90]{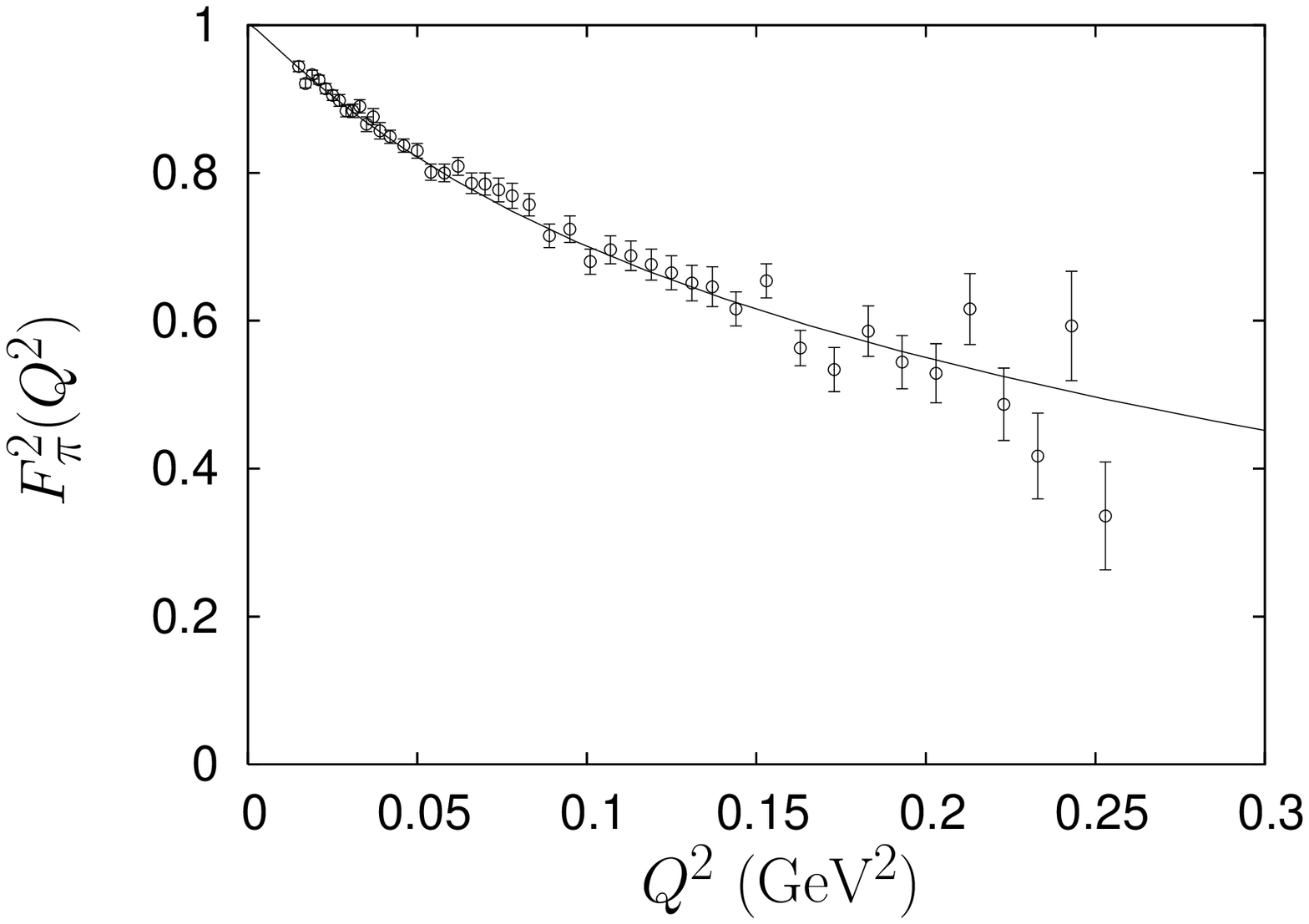}

\includegraphics[height=12cm,angle=-90]{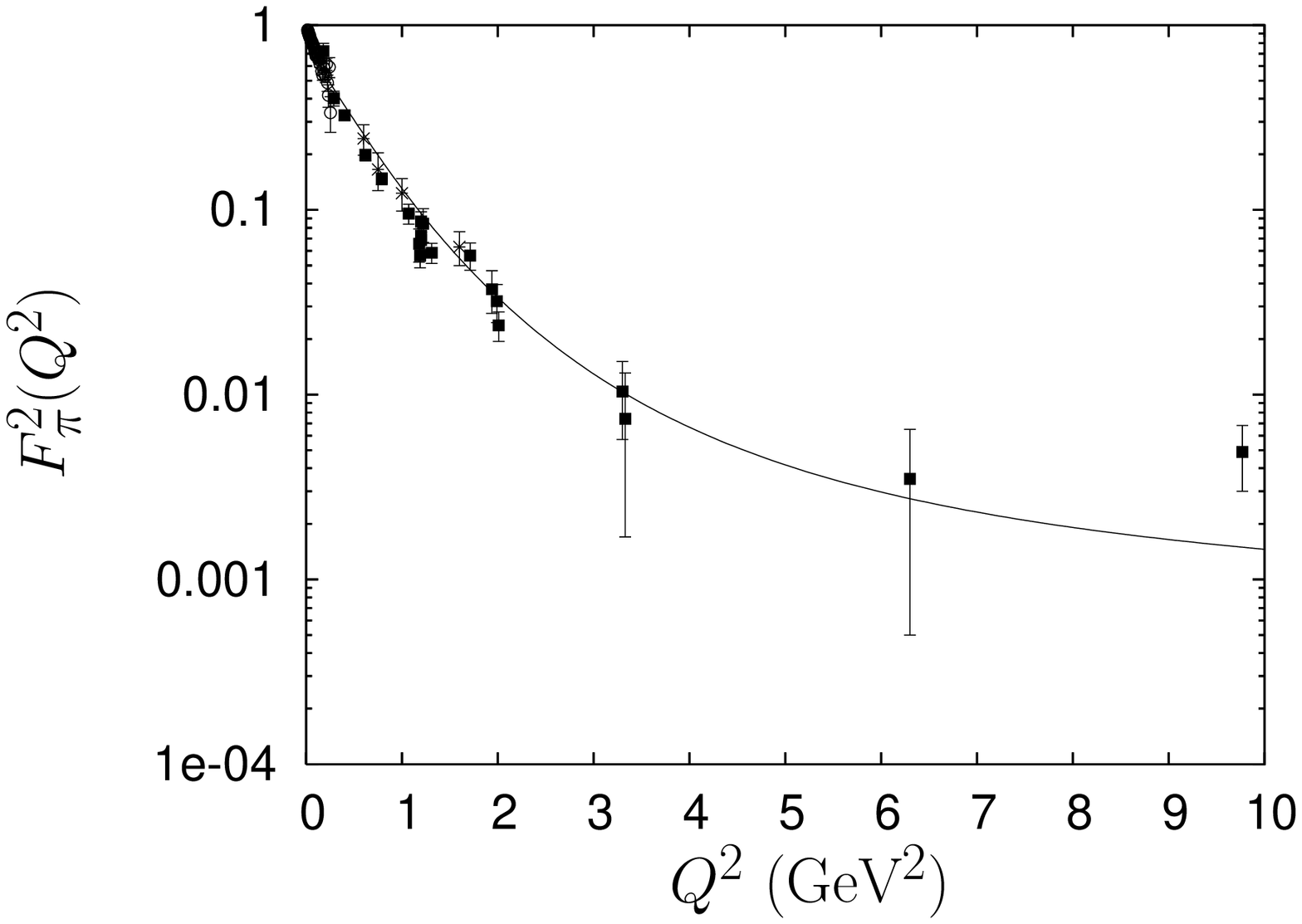}

  \caption{The charged pion form factor squared in comparison with
    experimental data from Refs.~\cite{amend} (open circles),
    \cite{bebek} (solid squares) and \cite{jlab} (crosses).}  
  \label{fig:fpi2}
\end{figure}

\begin{figure}
  \centering
  \includegraphics[height=12cm,angle=-90]{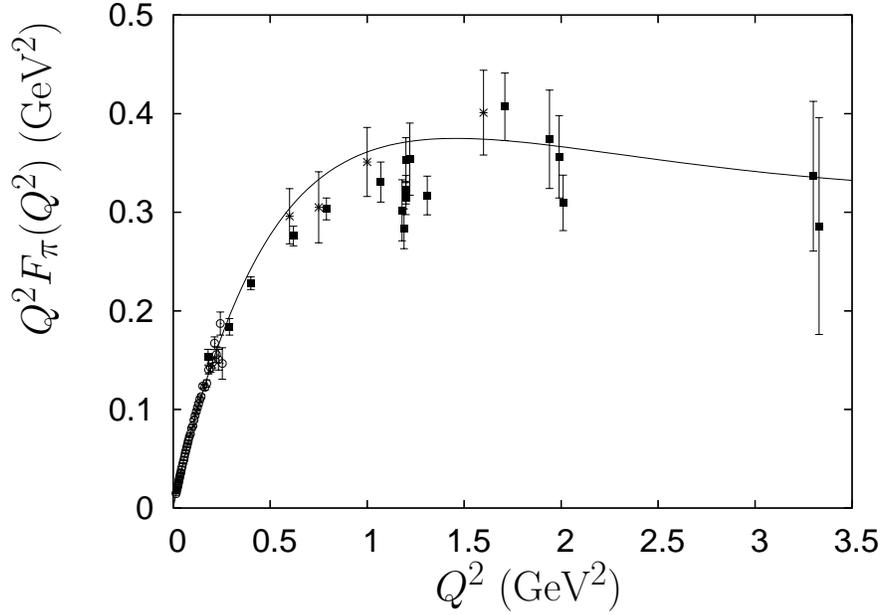}
  \caption{$Q^2$ times charged pion form factor in comparison with
    experimental data from Refs.~\cite{amend} (open circles),
    \cite{bebek} (solid squares) and \cite{jlab} (crosses). }
  \label{fig:q2fpi}
\end{figure}

\begin{figure}
  \centering
\includegraphics[height=12cm,angle=-90]{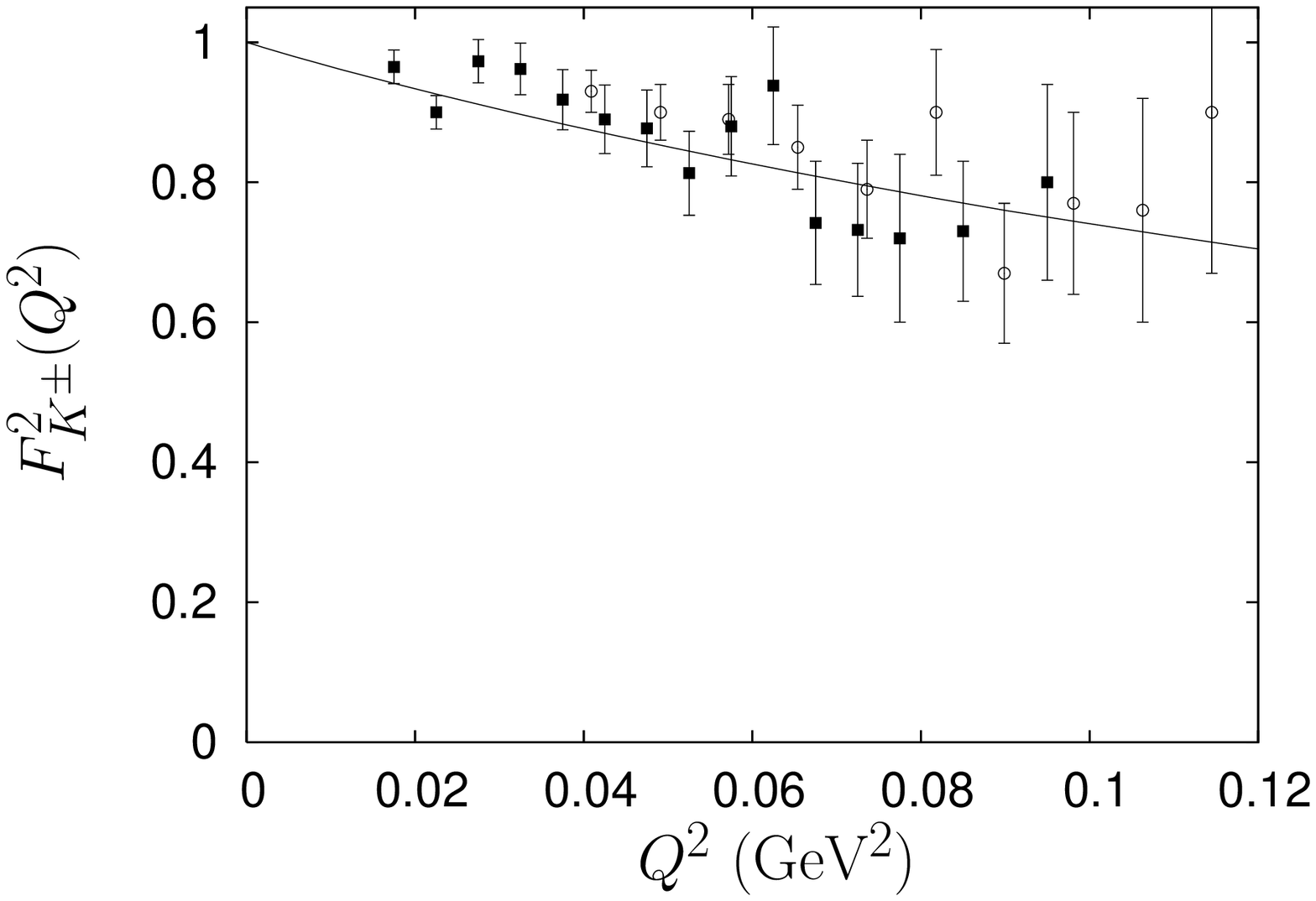}

  \caption{The charged kaon form factor squared in comparison with
    experimental data from Refs.~\cite{dally} (open circles) and
    \cite{amend2} (solid squares).}  
  \label{fig:fk2}
\end{figure}

\begin{figure}
  \centering
 \includegraphics[height=12cm,angle=-90]{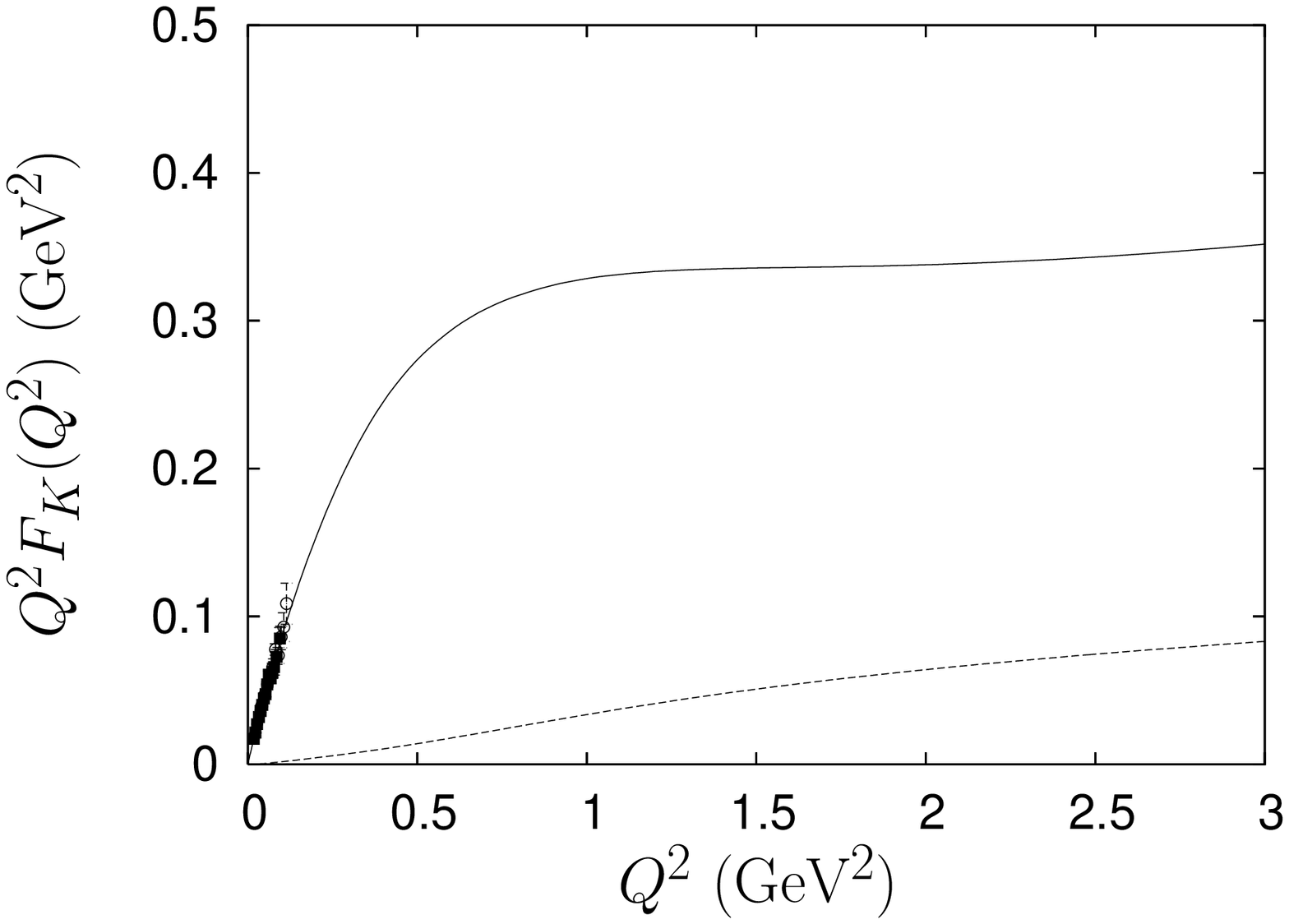}
  \caption{$Q^2$ times the charged kaon (solid line) and
    neutral kaon (dashed line) form factors. }
  \label{fig:q2fk}
\end{figure}

Now we can use the wave functions of the pseudoscalar light mesons
($\pi$, $K$), found in
Sec.~\ref{sec:lmm}, for the numerical calculation of their
electromagnetic form factors $F_P(Q^2)$ in the space-like region
$Q^2\ge 0$. The results of such calculations for the charged pion are shown
in Figs.~\ref{fig:fpi2} ($F_\pi^2(Q^2)$) and \ref{fig:q2fpi}
($Q^2F_\pi(Q^2)$) in comparison with 
experimental data from Refs.~\cite{amend,bebek,jlab}. Good agreement
with data both in low and high $Q^2$ regions is found, including recent
JLab data \cite{jlab} which are
plotted with crosses.  It is clearly seen
from Fig.~\ref{fig:q2fpi} that  the calculated pion form factor  at
high $Q^2$ exhibits the asymptotic behaviour $F_\pi(Q^2)\sim 
\alpha_s(Q^2)/Q^2$ predicted by the quark counting rule \cite{qcr}
and perturbative QCD \cite{asimpt}. Our results for the pion form
factor can also be compared with QCD based calculations
\cite{bpss} and with recent parameterizations  \cite{pp,bkk}
which arise  from the constraints of analyticity and unitarity. The
latter form factor models are based on the vector meson dominance and
include a 
pattern of radial excitations expected from dual resonance models
\cite{bkk}. The consistency of our results with such parameterizations 
(cf. Fig.~\ref{fig:fpi2} with Fig.~2 of Ref.~\cite{bkk}) just means
the manifestation of the quark-hadron duality. Finally, our predictions
agree fairly well with recent lattice computations of the pion form
factor~\cite{beflr,ha}.   
The corresponding plots for the charged kaon form factor are given in
Figs.~\ref{fig:fk2} and \ref{fig:q2fk} in comparison with experimental
data from Refs.~\cite{dally,amend2}, which are available only for the low
$Q^2$ region. Again  good agreement with experimental data is
found. On Fig.~\ref{fig:q2fk} we also plot the neutral kaon form
factor by the dashed line.

\begin{table}
  \caption{Charge radii of light pseudoscalar mesons.}
  \label{tab:cr}
\begin{ruledtabular}
\begin{tabular}{ccccccc}
charge radii& this work& \cite{gi}& \cite{mt} &\cite{hjd}&Lattice \cite{ha}
& Experiment \cite{pdg}\\ 
\hline
$\sqrt{\langle r^2\rangle_\pi}$ (fm) & 0.66&0.66&0.67&0.63&$0.63\pm0.1$  & 0.672$\pm$0.08 \\
$\sqrt{\langle r^2\rangle_{K^\pm}}$ (fm) & 0.57&0.59&0.62&0.60& & 0.560$\pm$0.031 \\
$\langle r^2\rangle_{K^0}$ (fm$^2$)& $-0.072$&$-0.09$&$-0.086$&$-0.062$&  & $-0.076\pm$0.018 
\end{tabular}
\end{ruledtabular}
\end{table}

The mean-squared charge radius of the pseudoscalar meson ($P=\pi,K$) is defined by
\begin{equation}
  \label{eq:chr}
  \langle r^2\rangle_P=-6\left[\frac{{\rm d}F_P(Q^2)}{{\rm d}Q^2}\right]_{Q^2=0}.
\end{equation}
The calculated values of the charge radii of light pseudoscalar mesons
are given in Table~\ref{tab:cr} in comparison with predictions of other
approaches \cite{gi,mt,hjd,ha} and experimental data \cite{pdg}. An
overall good agreement with experimental data is
found.

\section{Conclusions}
\label{sec:concl}

The relativistic quark model, which has been
previously developed and successfully used for the comprehensive
investigation of different properties of heavy and heavy-light
hadrons, was applied here for
calculating the masses, weak decay constants and
electromagnetic form factors of the light mesons. The main assumptions
and  parameters of the model (such as the Lorentz structure and
parameters of the confining 
potential and quark masses) were kept the same as in previous studies. The
only change we made, is the necessary modification of the running coupling
constant $\alpha_s(\mu^2)$ in the infrared region. 
Following Ref.~\cite{bvb} we chose
the simplest model with freezing (\ref{eq:alpha}). Therefore only one
additional parameter $\Lambda$ was introduced and it was fixed from
fitting the $\rho$ meson mass. We constructed the local relativistic
quasipotential for the light quarks using the replacement
(\ref{eq:sub}), which was previously tested on the heavy-light mesons.
The resulting relativistic potential (\ref{eq:v}) depends on the meson mass in a
complicated nonlinear way. Solving numerically the quasipotential
equation (\ref{quas}) we got masses of the ground-state and
radially-excited light mesons in a reasonably good overall agreement with
experimental data. Even the masses of the pseudoscalar $\pi$ and $K$
mesons are well reproduced. This is  a nontrivial result, since we use the
constituent quark masses in our description and thus the chiral
symmetry is explicitly broken from the very beginning. We determined
the light meson wave functions and used them for studying their
electroweak properties.           

First the weak decay constants of pseudoscalar and vector mesons were
investigated. It was argued that both positive- and negative-energy
parts of the quark propagators in the weak annihilation loop should be
taken into account. Usually in the semirelativistic quark model
\cite{gi,g,vd} only the positive-energy contributions are kept. This
approximation requires to replace in the expression for the
pseudoscalar decay constant (\ref{eq:fpv1}) the meson mass by
the so-called mock meson mass, which is considerably larger, in order
not to get the significant overestimate of the decay constants. We showed
that the negative-energy contributions to the light meson pseudoscalar
decay constants are large and negative. Their account brings
theoretical predictions (with the physical meson masses) in good agreement
with available experimental data.

Next we studied the electromagnetic form factor of the pseudoscalar
mesons. The corresponding matrix element of the electromagnetic
current was calculated using the quasipotential approach. The additional
contributions of the intermediate negative-energy states (\ref{gam2})
were taken into account as well as the transformation of the meson wave
function from the rest frame to a moving one (\ref{wig}). As a result
the relativistic expression for the electromagnetic form factor was
obtained. We then calculated the pion, charged and neutral kaon form factors
in the space-like region. Good agreement with available experimental
data both in small and large $Q^2$ regions were found. At large
momentum transfer this form factor tends to reproduce the power-law behaviour
predicted by perturtbative QCD \cite{asimpt}. The calculated charge
radii of light pseudoscalar mesons are in good agreement with
experiment.

In conclusion, we found that the obtained  results are
quite competitive with the predictions of other approaches
\cite{gi,mr,k,ak,mt,hjd,kt,milc,beflr,ha} including  more sophisticated ones,
which were specially developed for treating light mesons.     

\acknowledgments
The authors are grateful to A. Ali Khan, A. Badalian, M. M\"uller-Preussker,
V. Savrin and Yu. Simonov for useful discussions.  Two of us
(R.N.F. and V.O.G.)  were supported in part by the {\it Deutsche
Forschungsgemeinschaft} under contract Eb 139/2-3 and by the {\it Russian
Foundation for Basic Research} under Grant No.05-02-16243.

\end{document}